\documentclass[preprint,aps, preprintnumbers, nofootinbib]{revtex4}

\newcommand{\lsim}[1]{
\setlength{\unitlength}{12pt}
\begin{picture}(1.4,1.)
\put(.7,-0.3){\makebox(0.0,1.)[t]{$<$}}
\put(.7,-0.3){\makebox(0.0,1.)[b]{$\sim$}}
\end{picture}#1}
\newcommand{\gsim}[2]{
\setlength{\unitlength}{12pt}
\begin{picture}(1.4,1.)
\put(.7,-0.3){\makebox(0.0,1.)[t]{$>$}}
\put(.7,-0.3){\makebox(0.0,1.)[b]{$\sim$}}
\end{picture}#2}
\begin{document}

\title{Cosmic coincidences and relic neutrinos}


\author{R. Horvat\footnote{horvat@lei3.irb.hr}}
\affiliation{\footnotesize Rudjer Bo\v{s}kovi\'{c} Institute,
         P.O.B. 180, 10002 Zagreb, Croatia}


\begin{abstract}

A simple phenomenological description for the energy transfer between a
variable cosmological constant (CC) and a gas of relic neutrinos in an
expanding universe can account for  a near coincidence between the neutrino
and
dark-energy densities to hold over a significant portion of the history of
the
universe. Although such a cosmological setup may promote neutrinos to
mass-varying particles, both with slow and quick neutrino mass changing 
with the expansion of the universe naturally implemented in the model, it
also works equally well for static neutrino masses. We also stress what sort
of models for variable CC  can  potentially underpin the above
scenario.

\end{abstract}
\maketitle

Before 1997 there was a longstanding and hard-pressing problem in modern
physics called the cosmological constant (CC) problem \cite{1}. After a
remarkable discovery of dark energy starting in 1997 nothing spectacular
happened but the problem was dubbed the `old' CC problem. Besides subsequent
anthropic considerations, not much light has been shed since and the problem
remains practically intact. At the same time, by the discovery of dark
energy, at least two additional (and related) problems were also generated. In
the first one, one should explain why the CC is small but nonzero (now called
the `new' CC problem). In the second one (the `cosmic coincidence problem'),
one should explain the near coincidence of the CC energy density
($\rho_{\Lambda }$) towards the predominant component in the matter energy
density in the universe at present, i.e., cold dark matter 
($\rho_m $) \cite{2}. They are of the same order today, although vary in a 
completely different fashion over the history of the universe.

However, even the `cosmic coincidence problem' itself seems to be more profound 
since one
should also explain  today's coincidences of $\rho_{\Lambda }$ towards the
rest of the components of the universe: ordinary matter ($\rho_m $), 
radiation ($\rho_{\gamma }$), and
neutrinos ($\rho_{\nu }$). Indeed, although they all redshift quite differently, 
they all
become equal to $\rho_{\Lambda }$ within redshifts of order of a few. Now,
the bottom line is that, in distinction from other components, the past
cosmological behaviors of $\rho_{\Lambda }$ and $\rho_{\nu }$ are not much
known about. Concerning dark energy, all we know is that it is redshifting at
present with the equation of state (EOS) being very close to -1 \cite{3}. The
past cosmological behavior of $\rho_{\Lambda }$ could be to retain the same 
EOS in the past, thus creating a genuine `cosmic coincidence problem' at
present, or to track matter components by switching to its present EOS only in
the recent past, thus creating the `why now?' problem. However, this need not
necessarily be so for neutrinos. Namely, past and present cosmological
bounds on $\rho_{\nu }$ are quite weak, thus making a simple scaling
$\rho_{\Lambda } \sim \rho_{\nu }$ (within a factor of $10^2 -10^3$) the most
elegant resolution of the present coincidence between dark energy and
neutrinos.            

The first step towards  the above explanation for the near coincidence 
between
$\rho_{\Lambda }$ and $\rho_{\nu }$ was undertaken in \cite{4}. In the model 
\cite{4},
the dark sector included, besides the usual dark-energy sector given by a
quintessencelike scalar $\phi $, also a sector of relic neutrinos. The idea
was to promote the sector of relic neutrinos to an almost undilutable
quantity, as to be tightly bound with the $\phi $-sector. For that purpose,
since the neutrino number density $n_{\nu }$ 
in \cite{4} scales canonically, the
neutrino mass was  promoted to a running quantity 
\footnote{The idea of mass-varying neutrinos was first considered in
Ref.\cite{5}.}, thus scaling  almost as
$n_{\nu }^{-1}$. The dark sector, although being composed of two sectors,
acts as a single unified fluid.

Here, we present a far simpler model not including any extra dynamical  degree
of freedom (no quintessencelike scalars), in which dark energy
and neutrinos constitute separate sectors, but with energy being transferred
between them \cite{6}. The main ingredient is variable but `true' CC, with the EOS
being precisely -1. The continuous transfer of energy between the CC and the
gas of
relic neutrinos (and
{\it vice versa}, depending on the sign of the interaction term) can be
conveniently modeled by the generalized equation of continuity
\begin{equation}
\dot{\rho }_{\Lambda } + \dot{\rho }_{\nu } + 3H\rho_{\nu } (1+ \omega_{\nu
})
= 0 \;,
\end{equation}
where overdots denote time derivatives, and the EOS $\omega_{\nu}
$ for nonrelativistic neutrinos in (1), can be safely disregarded for all
practical purposes \footnote{Note that although at least two neutrino
species are strongly nonrelativistic today, a phase space distribution of
relic neutrinos still retains its relativistic form even for masses much
larger than those indicated by terrestial measurements, see, e. g. \cite{7}.
In addition, large neutrino mixing
revealed in neutrino oscillation experiments may serve to conclude that
chemical potentials for all neutrinos should be small \cite{8}.
Also, for the sake of simplicity, we
restrict ourselves here to the case of one neutrino family as in
\cite{4}; the generalization to three families of neutrinos is
straightforward.}. Now, let us make a specific {\it ansatz}
 for the energy transfer between the two components coupled through (1),
which leaves the number density of neutrinos to dilute canonically, but
promotes the mass of the neutrino to a running quantity, that is,
\begin{equation}
m_{\nu }(a)=m_{\nu_0 }a^{\alpha }; \;\;\;\;\;\;n_{\nu }(a)=n_{\nu_0 }
a^{-3}\;,
\end{equation}
where $\alpha $ is a constant, the subscript `0' denotes the present-day
values, and the present value for the scale factor is set to 1. The solution
for $\rho_{\Lambda }$ can be represented in a simple form
\begin{equation}
\rho_{\Lambda }(a) = \frac{\alpha }{3 - \alpha } \rho_{\nu }(a) +
\rho_{\Lambda
}^C\;,
\end{equation}
where both $\rho_{\nu }$ and $\rho_{\Lambda }$ now scale as
$ \sim a^{-3 + \alpha }$ and $\rho_{\Lambda }^C $ is the integration
constant. Note that $\rho_{\Lambda }^C $ represents the
true ground state of the vacuum. Regarding Eq. (5), several comments are in
order. If $\alpha > 0$, we are in the realm of decaying CC cosmologies
$(\dot{\rho}_{\Lambda } < 0)$, whereas $\alpha < 0$ means that the transfer
of energy is from neutrinos to the CC $(\dot{\rho}_{\Lambda } > 0)$.
Since the cosmic matter budget today
consists of no more than $2\%$ of massive neutrinos \cite{9}, one concludes
for the $\alpha > 0$ case that $\rho_{\Lambda }^C $ should always be
nonzero (and positive), unless $\alpha $ is fine-tuned to be very close to 3
(such large values of $\alpha $ are yet excluded, see below). On the other
hand, for $\alpha < 0$, the first term in (5) is negative, and hence also
the large and positive $\rho_{\Lambda }^C$ is required.

Now we try to restrict the $\alpha $-parameter by assuming the validity of
the generalized Second Law (GSL) of gravitational thermodynamics. The GSL
states that
the entropy of the event horizon plus the entropy of all the stuff in  
the volume inside the horizon cannot decrease in time. In a sense, it is
appropriate to do such an analysis here because we are dealing with cosmologies
in which ever accelerating universes always possess future event horizons.
The idea of associating entropy with the horizon area surrounding black holes is
now extended to include all event horizons \cite{10}. The easiest way to gain
information on the $\alpha $-parameter is by considering the change of
entropy in the asymptotic regime, $a \gg 1$. In this case, one should also
add the entropy of Hawking particles because it is conceivable that the CMB
temperature will drop below the Hawking temperature after some time in the
(distant) future \cite{11}. Thus we have 
\begin{eqnarray}
& \displaystyle\frac{d S_{tot}}{da}\geq 0 , & \nonumber \\& S_{tot}=S_{eh} +S_{de} + S_{m} +S_{\nu } + S_{\gamma } + S_{rgw} + S_{Hawk} , &
\end{eqnarray}
where the particular entropies in (4) are of the event horizon, dark energy, 
matter, neutrinos,
photons, relic gravitational waves, and of Hawking particles, respectively
\footnote{Note that although we are dealing with the `true' CC having
$w_{\Lambda } = -1$, its entropy
inside the event horizon $S_{de}$ does not vanish because our $\Lambda $ is a
varying quantity. Furthermore, $S_{rgw}$ is about to vanish at some time in
the distant future, see, e.g. \cite{12}. For details of such a type of
calculation as well as proper definitions for particular entropies, 
see \cite{18}.}. What we find is that for $\alpha < 0$ it is impossible to
satisfy the GSL, whereas for $\alpha > 0$ it is possible, provided that 
\begin{eqnarray}
& \displaystyle\alpha \lsim 3/4 , & \nonumber \\
& m_{\nu } > \frac{\sqrt{\rho_{\Lambda}^C }}{M_{Pl}}.   &
\end{eqnarray}
Note that the second restriction is trivially satisfied for nonrelativistic
neutrinos. Hence, the GSL prefers positive and less-than-one $\alpha $'s.

Further restriction on the $\alpha $-parameter can be obtained by using the
concept of effective EOS (for dark energy), as put forward by Linder and
Jenkins \cite{13}. The effective EOS for the variable CC whose interaction
is phrased by (1)
can be defined similarly as in \cite {13}
\begin{equation}
\omega_{dark}^{eff} = -1 + \frac{1}{3} \frac{d \; ln \; \delta H^2 (z)}
{d \; ln \; (1 +z)}\;,
\end{equation}
where $1+z = a^{-1}$. Here, any modification of the standard Hubble parameter
$H$ is encapsulated in the term $\delta H^2 $ (including $\rho_{\Lambda }^C
$).
For the model under consideration, one obtains
\begin{equation}
\omega_{dark}^{eff} = -1 + \frac{(1+z)^{3-\alpha } - (1+z)^3}{\left
(\frac{3}{3-\alpha } \right )(1+z)^{3-\alpha } - (1+z)^3 + \rho_{\Lambda
}^C/\rho_{\nu 0}} \;.
\end{equation}
Even though $\alpha $'s as large as $\gsim \; 1$ are sustained by (7)
because
the ratio $\rho_{\Lambda }^C /\rho_{\nu 0}$ can be large, such large
values for $\alpha $ would spoil the tracking behavior at earlier times
when the constant term in the denominator of (7) ceased to be  dominant,
and therefore the only acceptable values are $\alpha \ll 1$. By combining
this arguments with the arguments from the GSL, one sees that slow mass
variation over cosmological scales is preferable. 

Note that with another {\it ansatz}, in which the total number of neutrinos in a
comoving volume changes while retaining its proper mass constant,
\begin{equation}
m_{\nu }(a)=m_{\nu_0 }; \;\;\;\;\;\;n_{\nu }(a)=n_{\nu_0 }
a^{-3 +\beta }\;,
\end{equation}
with $\beta $ being a constant, the `coincidence' 
$\rho_{\Lambda } \sim \rho_{\nu }$ is still maintained [simply make the 
replacement $\alpha \rightarrow \beta $ in (3)]. The nice feature that
$\rho_{\Lambda }$, as a solution of (1), always tracks $\rho_{\nu }$, is
maintained even when both the total number of neutrinos and their proper
masses
change, i.e.,
\begin{equation}
m_{\nu }(a)=m_{\nu_0 }a^{\alpha }; \;\;\;\;\;\;n_{\nu }(a)=n_{\nu_0 }
a^{-3 +\beta }\;.
\end{equation}    
In this case, one can easily check that the effective EOS is still given by
(7), but now
with the replacement $\alpha \rightarrow \alpha + \beta$, and therefore the
requirement $\alpha << 1$ turns into $\alpha + \beta \approx 0$, signaling
thus a two-way energy transfer between $\Lambda $ and relic neutrinos. In
this way we obtain quick neutrino mass changing with the expansion of the
universe.  

We would like to mention that the variable CC model \cite{14} is
completely able to underpin the present scenario. It is a decaying CC model
with $\alpha + \beta > 0$. The model is based on the renormalization-group
(RG) evolution for $\rho_{\Lambda }$, and on the choice for the RG scale
$\mu =
H$. It can be shown that a canonical value for $\alpha + \beta $ is $(4\pi
)^{-1} \ll 1$. In addition, the CC-variation  law,
$d\rho_{\Lambda }/dz \propto
dH^2/dz$, is a derivative one, thus having a natural appearance of a
nonzero $\rho_{\Lambda }^C$. Finally, we consider a dynamical CC scenario
generically dubbed `holographic dark energy' (HDE).
Derived originally for zero-point energies, the
Cohen {\it et al.} bound \cite{15} for $\rho_{\Lambda }$ can be rewritten
in the form
\begin{equation}
\rho_{\Lambda}(\mu)=\kappa \mu^2 G_N^{-1},
\end{equation}
where $\mu $ denotes the IR cutoff and $\kappa $ represents a degree of
saturation  of the bound. This is a very important concept since for natural
values for $\kappa $ of order of unity, the HDE model
also represents  one of the most elegant solution of the `old' CC problem.
Through the relationship between the UV ($\rho_{\Lambda } \sim {\Lambda }^4
$) and the IR cutoff as given by (10), the holographic information is
consistently encoded in ordinary quantum field theory. The most natural and
simplest possibility is to have $\mu = H$. In our case, however, $\mu \sim H$
unavoidably implies $\rho_{\Lambda }^C = 0$, thus spoiling the successfulness of
the scenario \footnote{Other popular choices for $\mu $ like the inverse
particle horizon or inverse future horizon also lead to an unrealistic
scenario.}. Still, agreement is possible for noncanonical choices for $\mu $
or modification of the law (10), for instance, by promoting $\kappa $
\cite{16} or $G_N $ \cite{17} to a varying quantity.  

Let us summarize our main results and conclude by a few additional comments.
The variable part of $\rho_{\Lambda }$ dilutes at the same rate as $\nu $'s
and hence we have $\rho_{\Lambda } \sim \rho_{\nu }$ for a large portion of
the history of the universe. One the other hand, the tracking behavior
without the constant part in $\rho_{\Lambda }$ leads to an unrealistic model
of the universe. For $\alpha << 1$, we also have an approximate
tracking of $\rho_{\Lambda }$, $\rho_{\nu }$ and $\rho_m \sim a^{-3}$,
because
in our scenario matter dilutes canonically. Note that the positivity of $\alpha
$ is essential here to have dominance of matter over neutrinos in the past. 
If the interaction between
$\Lambda $ and $\nu $'s is maintained in the epoch where $\nu $'s become
relativistic, one can be easily convinced, considering  a replacement 
$-3 + \alpha
\rightarrow -4 + \alpha $ in the above expressions, that then we have an
approximate tracking of $\rho_{\Lambda }$, $\rho_{\nu }$ and $\rho_{\gamma }
\sim a^{-4}$. The scenario is generally underpinned by renormalization-group 
running cosmologies. In its simplest and originally derived form, the 
scenario is not generally underpinned by HDE, but fits the generalized 
HDE models.

\end{document}